\documentclass[reprint,superscriptaddress,amsmath,amssymb,aps]{revtex4-1}
\usepackage[utf8]{inputenc}
\usepackage[T1]{fontenc} 
\usepackage{graphicx}
\usepackage{rotating}
\usepackage{makeidx}
\usepackage{threeparttable}
\usepackage{float}
\usepackage{amsmath}
\usepackage{rotating}
\usepackage{hyperref}

\usepackage{relsize}
\usepackage{etoolbox}
\usepackage[most]{tcolorbox}
\usepackage{hyperref}
\usepackage{dcolumn}
\usepackage{bm}
\usepackage{wrapfig}
\usepackage{float}  
\usepackage[toc,page]{appendix}
\usepackage{gensymb}
\usepackage{epstopdf}
\usepackage{subfiles}
\usepackage{comment} 
\usepackage[autostyle=true]{csquotes} 
\usepackage{makecell}
\usepackage{siunitx}
\usepackage[all]{hypcap} 
\usepackage[most]{tcolorbox}
\usepackage{hyperref}
\usepackage{xcolor}
\usepackage{nicefrac}
\usepackage{chemformula}
\usepackage{ulem}  %\sout
\definecolor{gainsboro}{rgb}{0.86, 0.86, 0.86}
\usepackage[backgroundcolor=gainsboro]{todonotes}
\hypersetup{
  colorlinks   = true, %Colours links instead of ugly boxesr
  urlcolor     = blue, %Colour for external hyperlinks
  linkcolor    = blue, %Colour of internal links
  citecolor   = blue %Colour of citations
}

\newcommand{\weg}[1]{}

\newcommand{\obyth}[2]{$\frac{1}{3}$}
\newcommand{\obyf}[2]{$\frac{1}{4}$}

\begin{document}

\title{Photoinduced orbital polarization and Jahn-Teller effect in RNiO$_3$}

\author{Sangeeta Rajpurohit}
\email{rajpurohit1@llnl.gov}
\affiliation{Material Science Division, Lawrence Livermore National Laboratory, CA 94550, USA}
\author{Sheikh Rubaiat Ul Haque}
\affiliation{Department of Applied Physics, Stanford University, Stanford, CA 94305, USA}
\affiliation{Department of Materials Science and Engineering, Stanford University, Stanford, CA 94305, USA}
\affiliation{SLAC National Accelerator Laboratory, Menlo Park, CA 94025, USA}
\author{Aaron M. Lindenberg} 
\affiliation{Department of Materials Science and Engineering, Stanford University, Stanford, CA 94305, USA}
\affiliation{Stanford Institute for Materials and Energy Sciences, SLAC National Accelerator Laboratory, Menlo Park, CA 94025, USA}

\author{Peter E. Blöchl}
\affiliation{Clausthal University of Technology, Institute of Theoretical Physics, 38678 Clausthal-Zellerfeld, Germany}
\affiliation{University of Göttingen, Institute of Theoretical Physics, Friedrich-Hund-Platz 1, 37077 Göttingen, Germany}
\author{Tadashi Ogitsu}
\affiliation{Material Science Division, Lawrence Livermore National Laboratory, CA 94550, USA}

\date{\today}

\begin{abstract}

The orbital degree of freedom in rare-earth nickelates is typically inactive across the temperature-driven
metal-insulator transition, where the system develops two inequivalent Ni sites associated with Ni-O bond
disproportionation and breathing-mode distortions of NiO$_6$ octahedra. Here, we show that orbital polarization
can be induced by optical excitation with linearly polarized light. Using an interacting multiband tight-binding model
combined with real-time simulations of coupled electron-ion-spin dynamics, we find that photoinduced $d$-$d$
transitions reduce the local magnetic moments at Ni sites and effectively suppress Hund’s coupling $J$ in the
excited state. Importantly, these transitions can be made strongly orbital-selective by tuning the light polarization,
leading to an imbalance in $e_g$ orbital occupancies. The resulting nonequilibrium state, characterized by reduced
effective $J$ and unequal orbital populations, becomes unstable toward Jahn-Teller (JT) distortions, driving
structural relaxation along coherently excited JT modes. Our results demonstrate that polarization-controlled
optical excitation provides a pathway to access hidden nonthermal phases with emergent orbital order, enabling
coherent control of coupled charge, spin, and lattice degrees of freedom on ultrafast timescales.
\end{abstract}

%provides a nonthermal pathway to activate orbital dynamics.

\maketitle

\noindent
\textit{Introduction:} Rare-earth nickelates RNiO$_3$ (R $\neq$ La) exhibit a temperature-driven metal-insulator transition (MIT)
into a magnetically ordered insulating phase \cite{Alonso1999,Alonso2001,Medarde1997}. Unlike conventional Mott insulators,
the low-temperature phase is better described by bond disproportionation, where NiO$_6$ octahedra alternately expand and
contract along all three directions, lowering the symmetry from orthorhombic $Pbnm$ to monoclinic $P2_1/n$. In this picture,
charge disproportionation (CD) is often described as alternating Ni$^{3-\delta}$ and Ni$^{3+\delta}$ sites.
An alternative description invokes a negative charge-transfer scenario, with Ni$^{2+}$ ($d^8$) and Ni$^{2+}$ ($d^8\underline{L}^2$) configurations,
where $\underline{L}$ denotes a ligand hole \cite{Mizokawa2000_2,Varignon2017}. Across the RNiO$_3$ series, the MIT and Néel
temperatures vary strongly with the rare-earth ionic size, which controls the bandwidth. In higher-bandwidth compounds, the MIT
coincides with magnetic ordering, whereas in lower-bandwidth systems, antiferromagnetic order appears at a lower temperature.
For the latter, a site-selective Mott transition has been proposed: $d$ electrons on Ni$^{2+}$ ($d^8$) sites form localized moments,
while those on Ni$^{2+}$ ($d^8\underline{L}^2$) sites form singlets with ligand holes \cite{Park2012}.

Despite this rich interplay of charge, spin, and lattice degrees of freedom, the orbital degree of freedom in bulk RNiO$_3$ remains
largely inactive, with no clear evidence of symmetry-breaking orbital order. While strain \cite{Stewart2011,Chakhalian2011},
doping \cite{Torrance1992}, and electromagnetic fields \cite{Nicoletti2016,Caviglia2013,Stoica2022} can tune these properties,
they have not enabled deterministic control of orbital polarization. Some studies report orbital-order-driven MITs under
strain \cite{Peil2014,He2015}, but a clear route remains elusive.
The entagled nature of charge, spin, lattice, and orbital degrees of freedom makes their individual roles difficult to isolate
in equilibrium. Ultrafast pump-probe techniques perturb these couplings on intrinsic timescales, enabling selective control.
For example, time-resolved x-ray diffraction and optical pump-probe studies on NdNiO$_3$ show that photoexcitation
quenches magnetic order on sub-100 fs timescales, followed by a slower collapse of bond disproportionation over several
hundred femtoseconds \cite{Caviglia2013,Stoica2022}. However, the possibility of using light to control local $d$-orbital
polarization at Ni sites and potentially induce long-range orbital order remains largely unexplored. In this work, we demonstrate
that optical excitation provides a powerful route to transiently lift the Ni $d$-orbital degeneracy in RNiO$_3$. 

We employ a multi-orbital, interacting tight-binding (TB) model that captures the low-energy physics of Ni $e_g$-electrons
in RNiO$_3$. A relatively small $(U{-}3J)$, where $U$ and $J$ denote the on-site Coulomb interaction and Hund’s exchange
coupling, respectively, favors an insulating ground state with charge disproportionation of the type $2\mathrm{Ni}^{3+} \rightarrow \mathrm{Ni}^{4+} + \mathrm{Ni}^{2+}$.
In this state, the $\mathrm{Ni}^{2+}$ sites adopt a high-spin configuration, and the insulating phase is further stabilized by breathing-mode
distortions of the NiO$_6$ octahedra. Our real-time simulations show that, under optical excitation with linearly polarized light, $d$-$d$
transitions drive a spin-unpolarized charge transfer from Ni$^{2+}$ to Ni$^{4+}$ sites. This process reduces the effective Hund’s coupling $J$,
thereby enhancing $(U{-}3J)$. By tuning the light polarization, the charge transfer can be made orbital-selective, leading to an imbalance in $e_g$
orbital occupations. The resulting photoinduced state, characterized by pronounced orbital polarization, becomes unstable toward
Jahn-Teller (JT) distortions. The explicit inclusion of coupled charge, spin, and lattice dynamics in our time-dependent simulations enables
us to disentangle their respective roles in the emergence, lifetime, and decay of this hidden orbital-polarized state.

\noindent
\textit{Model:} In the octahedral field of oxygen, the Ni--$3d$ shell splits into a completely-filled triply degenerate $t_{2g}$-set and a partially filled
$e_g$-doublet. The $e_g$-electrons delocalize
across neighboring Ni sites via oxygen-bridged hopping and are treated quantum mechanically
in our TB-model. The total potential energy of the system in our model is
\begin{equation}
E_{\mathrm{pot}} \;=\; E_e \;+\; E_{\mathrm{ph}} \;+\; E_{e\text{-}\mathrm{ph}},
\label{eq:tbm_1}
\end{equation}
with $E_e{=}E_{\mathrm{hop}} {+} E_{\mathrm{coul}}$. The $e_g$-electrons are described by
single-particle Pauli spinors $
|\psi_n(t)\rangle{=} \sum_{R}\sum_{\sigma}\sum_{\alpha\in\{a,b\}}
|\chi_{\sigma,\alpha,R}\rangle \, \psi_{\sigma,\alpha,R,n}(t),
$
where $n$ is the band index with $f_n$ as their corresponding occupations. The basis set $\{|\chi_{\sigma,\alpha,R}\rangle\}$ consisting of Ni-centered $e_g$-like
orbitals pointing toward the oxygen ligands. The spin index is
$\sigma\in\{\uparrow,\downarrow\}$. The orbital index is $\alpha\in\{a,b\}$
with $a=d_{x^2-y^2}$ and $b=d_{3z^2-r^2}$. The site index $R$
runs over Ni sites. Oxygen character is included by downfolding, so these
$e_g$-orbitals are antibonding. 
The hopping energy is
\begin{equation}
E_{\mathrm{hop}} \;=\; 
\sum_{R,R',\sigma,n} f_n
\sum_{\alpha,\alpha'}
\psi^{*}_{\sigma,\alpha,R,n}\,
t_{\alpha\alpha'}(R,R')\,
\psi_{\sigma,\alpha',R',n},
\label{eq:tbm_2}
\end{equation}
where $t_{\alpha\alpha'}(R,R')$ are hopping matrix elements
between Ni neighbors linked by an oxygen bridge. The on-site Coulomb interactions at the mean-field level within the $e_g$ subspace are
\begin{eqnarray} 
&E&^R_{coul}= \frac{U}{2}\sum_{\substack{\sigma\ne\sigma' \\\alpha}}n_{\sigma\alpha,R}n_{\sigma'\alpha,R} {+}\frac{U{-}3J}{2}\sum_{\substack{\sigma\\\alpha\ne\alpha'}}n_{\sigma\alpha,R}n_{\sigma\alpha',R} \nonumber \\ &+&\frac{U{-}2J}{2}\sum_{\substack{\sigma\ne\sigma'\\\alpha\ne\alpha'}}n_{\sigma\alpha,R}n_{\sigma'\alpha',R} {-}\frac{J}{2}\sum_{\substack{\sigma\ne\sigma'\\\alpha\ne\alpha'}}\big(\rho_{\sigma,\alpha,\sigma',\alpha,R}\rho_{\sigma',\alpha',\sigma,\alpha',R} \nonumber \\ &{+}& \rho_{\sigma,\alpha,\sigma,\alpha',R}\rho_{\sigma',\alpha,\sigma',\alpha',R}\big) 
\label{eq:tbm_2} 
\end{eqnarray}
with the on-site one-body density matrix
\begin{equation} 
\rho_{\sigma,\alpha,\sigma',\alpha',R}{=}\sum\limits_{n} f_n\psi_{\sigma,\alpha,R,n}\psi^*_{\sigma',\alpha',R,n}. 
\label{eq:densmat} 
\end{equation}
Equation~\eqref{eq:tbm_2} has the standard two-orbital Kanamori form
\cite{Junjiro1963,Jernej2013}. The first three terms describe opposite-spin
repulsion in the same orbital, same-spin repulsion in different orbitals,
and opposite-spin repulsion in different orbitals. The last two terms in Eqn. \ref{eq:tbm_2} are
spin-flip and pair-hopping. 
The el-ph coupling and lattice energy terms are
\begin{eqnarray}
E_{e\text{-}\mathrm{ph}} &=&
g_{\mathrm{JT}}
\sum_{R,\sigma}\sum_{\alpha,\beta}
\rho_{\sigma\alpha,\sigma\beta,R}\,
M^{Q}_{\beta\alpha}\!\big(Q_{1,R},Q_{2,R},Q_{3,R}\big), \\
E_{\mathrm{ph}} &=&
\frac{1}{2}k_{\mathrm{JT}}\sum_{R}
\left(
Q_{2,R}^{2}+Q_{3,R}^{2}
+\frac{k_{\mathrm{br}}}{k_{\mathrm{JT}}}\,Q_{1,R}^{2}
\right).
\label{eq:eq_el-ph}
\end{eqnarray}
with
\begin{equation}
\mathbf{M}^{Q}(Q_{1,R},Q_{2,R},Q_{3,R}) \;=\;
\begin{pmatrix}
Q_{3,R} & Q_{2,R}\\[2pt]
Q_{2,R} & -Q_{3,R}
\end{pmatrix}
\;-\;
\mathbf{1}\,\frac{g_{\mathrm{br}}}{g_{\mathrm{JT}}}\,Q_{1,R}.
\end{equation}
describe the coupling between local charge density and
eg-orbital polarization with the breathing mode $Q_{1}$ and
the two JT modes $Q_{2/3}$, respectively. 
Here, $g_{\mathrm{JT}}$ and $g_{\mathrm{br}}$ are the JT and
breathing-mode el-ph couplings. We take the model parameters reported in \cite{Sotoudeh2016}, which were systematically
derived from first-principles calculations. Specifically, we use $t_{\mathrm{hopp}}{=}0.45$ eV, $U{=}2.0$ eV,
$U/J{=}2.5$, $k_{\mathrm{br}}{=}12.04$ eV/\AA$^2$,$k_{\mathrm{br}}/k_{\mathrm{JT}}${=}2.39, $g_{\mathrm{JT}}{=}2.50$ $eV$/\AA,
and $g_{\mathrm{br}}{=}2.0$ $eV$/\AA.

\begin{figure}[!thp]
\begin{center}
\includegraphics[width=\linewidth]{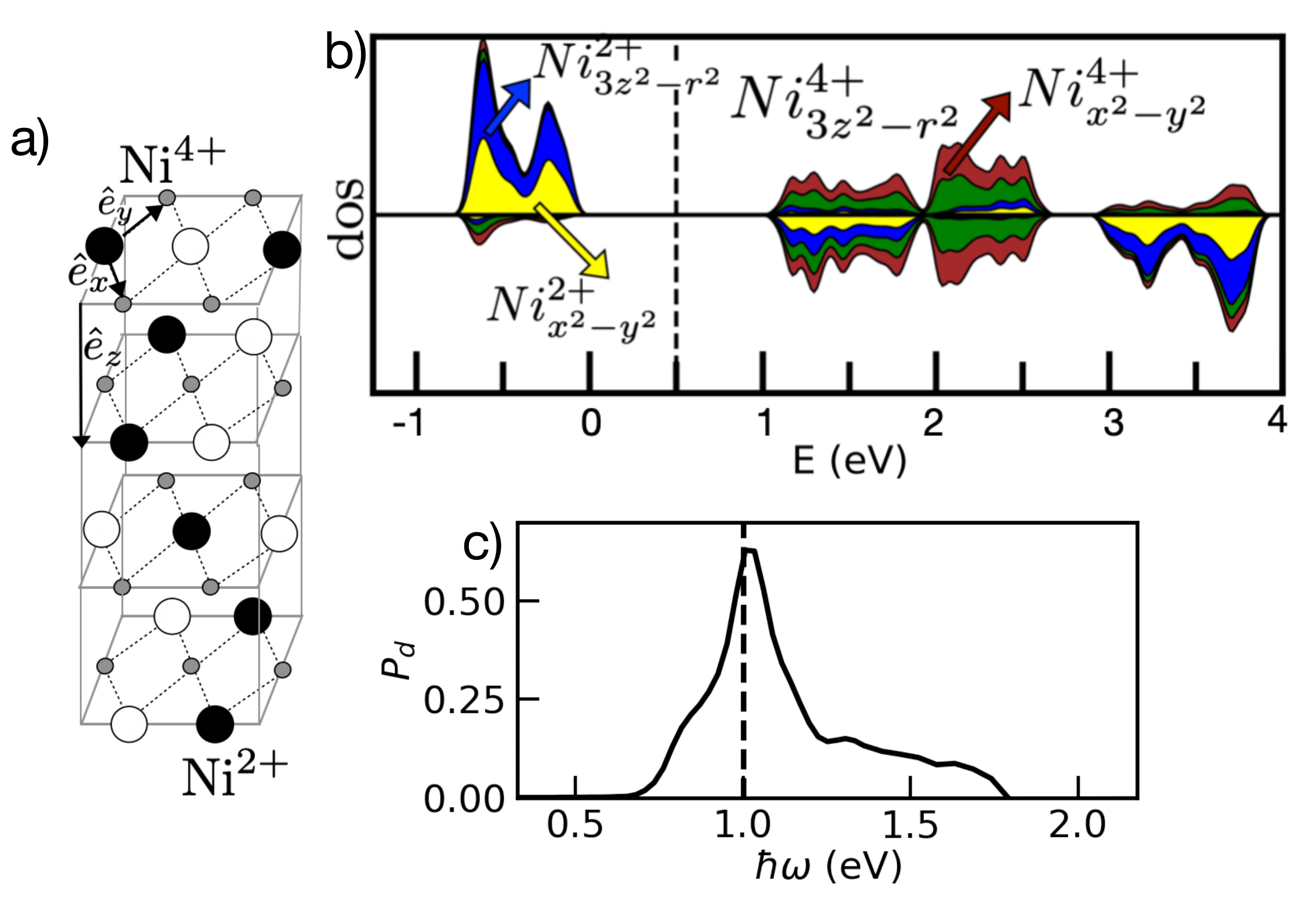}
\end{center}
\vspace{-0.5 cm}
\caption{\textbf{(a)} Schematic of spin and charge order in the E-type 
AFM insulating phase. Ni$^{2+}$ (Ni$_L$) and Ni$^{4+}$ (Ni$_S$) sites
are indicated by larger and smaller circles, respectively. Black/white
fills denote up/down moments on Ni$_L$ sites; Ni$_S$ sites are nonmagnetic
in the ground state. \textbf{(b)} Projected density of states (PDOS) of the
AFM phase onto the two $e_g$ orbitals at Ni$_L$ and Ni$_S$ sites.
\textbf{(c)} Photo-absorption spectrum of RNiO$_3$ as a function of
photon energy. Vertical axes denotes photon-absorption density $P_d$
and horizontal axes photon energy $\hbar\omega$. The dashed vertical
line denotes $\hbar\omega_o=$1.0 eV, which is the energy adopted in the 
subsequent simulations. }
\label{fig:fig1}
\end{figure}

\noindent
\textit{Ground-state:}
For the chosen parameters, the model stabilizes an antiferromagnetic insulating state with charge disproportionation (CD) between inequivalent Ni sites, 
$2\mathrm{Ni}^{3+} \rightarrow \mathrm{Ni}^{4+} + \mathrm{Ni}^{2+}$ as shown in Fig.~\ref{fig:fig1}(a). The $\mathrm{Ni}_L$ sites
carry larger local moments, consistent with a high-spin configuration. This is consistent with the experimentally observed
low-temperature phase of RNiO$_3$ below $T{\mathrm N}$ \cite{Alonso1999}. Breathing-mode distortions of the NiO$_6$ octahedra
further stabilize the CD state, forming a three-dimensional bond-disproportionated structure with alternating expanded and compressed
octahedra around $\mathrm{Ni}_L$ and $\mathrm{Ni}_S$ sites, respectively. The spin-resolved $e_g$-projected density of states
[Fig.~\ref{fig:fig1}(b)] shows that occupied states near the Fermi level are predominantly localized on $\mathrm{Ni}_L$ sites, while
 low-energy unoccupied states are associated with $\mathrm{Ni}_S$ sites, reflecting the underlying charge and spin asymmetry.

%########################################################
%####################FIGURE START########################
%########################################################
%########################################################
%####################FIGURE ENDS########################
%########################################################

For small $(U{-}3J)$, relevant to RNiO$_3$ physics, the system favors a charge-disproportionation-driven insulating phase.
Increasing $(U{-}3J)$ drives the system toward an instability associated with JT distortions and orbitally polarized states, as
 observed in other classes of correlated oxides such as PrMnO$_3$ \cite{Sotoudeh2016,Mazin2007}.

In this work, we demonstrate that orbital-selective optical excitation provides a nonthermal route to dynamically reduce Hund’s
coupling $J$, thereby increasing $(U{-}3J)$. In the photoexcited state, an imbalance in $e_g$ orbital populations
($\Delta n^{eg}_R \neq 0$) lowers the electronic energy sufficiently to overcome the elastic cost of JT distortions. This drives
the system toward a homogeneous, orbitally polarized metastable state in which charge disproportionation is suppressed.
  
\noindent
\textit{Real-time dynamics of photoexcited RNiO$_3$:}
We simulate the coupled electron-ion-spin dynamics under optical excitation using Ehrenfest dynamics. The electronic subsystem
evolves according to the time-dependent Schrödinger equation, while the nuclei are propagated classically via Newton’s equations
of motion. The light pulse is modeled as a spatially homogeneous time-dependent field introduced through the vector potential
$\mathbf{A}(t)=\mathbf{A}_0g(t)$, with a Gaussian envelope $g(t)=\exp(-t^{2}/2c_w)$, and incorporated via the Peierls substitution~\cite{Hofstadter1976}.

We consider a $4\times4\times4$ supercell of Ni sites with a $4\times4\times4$ $\Gamma$-centered $k$-grid. The initial state is the
E-type AFM insulating phase [Fig.~\ref{fig:fig1}(a)]. A 75-fs pulse is applied, and the response is studied for two polarization directions,
$\hat{\mathbf{e}}{x}{+}\hat{\mathbf{e}}{y}{+}\hat{\mathbf{e}}{z}$ and $\hat{\mathbf{e}}{z}$, where $\hat{\mathbf{e}}_{i}$ are along the Ni-O bonds.

\textit{Dipole-allowed transitions:}
Optical excitations in RNiO$3$ involve both $d$-$d$ and $p$-$d$ transitions in the energy range $\sim$0.5-6 eV \cite{Moskvin2010,Ifland2017}.
Low-energy ($\lesssim$2 eV) absorption is dominated by dipole-allowed transitions within the $e_g$ manifold, while higher-energy features
arise from $t_{2g}$ and O-$2p$ to $e_g$ excitations \cite{Stewart2011,Torriss2017}. Here we focus on $e_g \rightarrow e_g$ transitions. 
The calculated absorption spectrum [Fig.~\ref{fig:fig1}(c)] peaks at $\hbar\omega_0{=}1.0$ eV, which we use in the following simulations.

\begin{figure}[tp!]
     \begin{center}
     \includegraphics[width=1.0\linewidth]{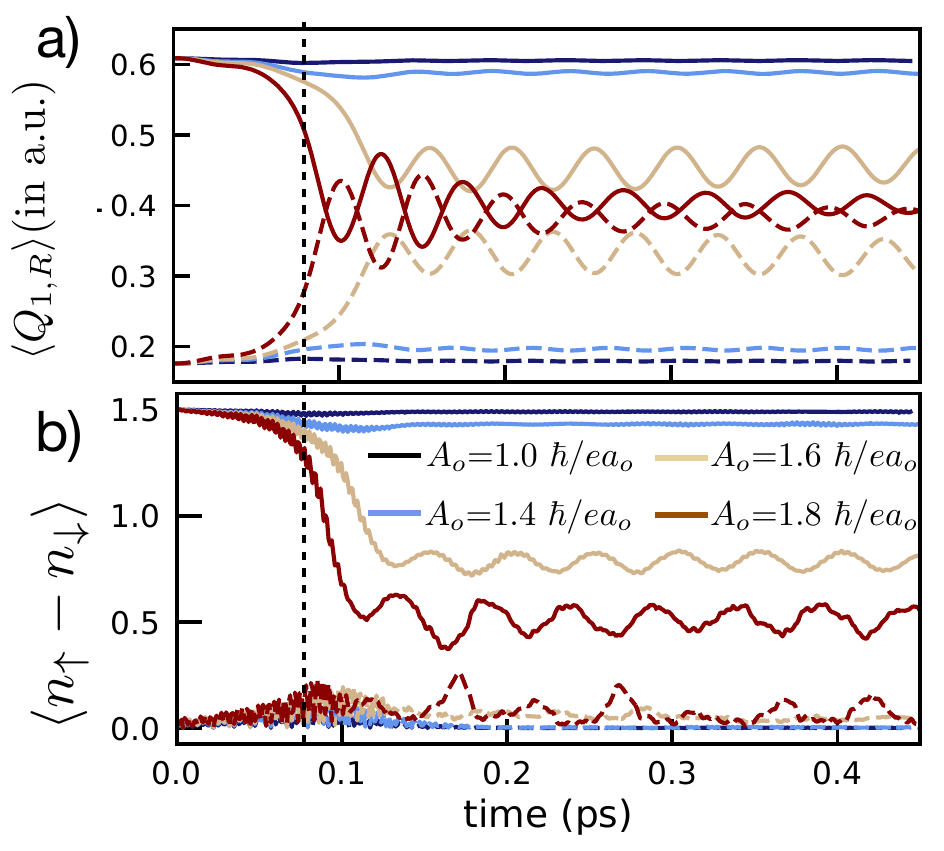}
      \caption{Time evolution of the breathing-mode distortion \(Q_1\) \textbf{(a)} and local magnetic moment distortion \textbf{(b)} at
       Ni\(_L\) (solid) and Ni\(_S\) (dashed) sites for different pump fluences, during and after photoexcitation with light polarized
     along \(\hat{\mathbf e}_z\). The dashed vertical lines show the center of the 75 fs Gaussian-shaped pulse.}
     \label{fig:fig2}
     \end{center}
\end{figure}

\textit{Charge and spin dynamics:} Figure \ref{fig:fig2}(a) and (b) shows the time evolution of the site-resolved breathing modes
$Q_{1,R}$ and local magnetic moments for different pulse amplitudes. Photoexcitation promotes electrons from occupied $e_g$ states
on Ni$_L$ to unoccupied $e_g$ states on Ni$_S$, driving charge transfer from $\mathrm{Ni}_L$ to $\mathrm{Ni}_S$ and partially
suppressing the charge disproportionation. Because the local charge density is strongly coupled to the breathing mode $Q_{1,R}$,
this modification of the charge disproportionation is reflected in the oscillations of $Q_{1,R}$ (see Fig.~\ref{fig:fig2}(a)), which remain coherent over
0.5 ps timescales with negligible damping within our simulation window. The suppression of charge disproportionation increases with field strength,
and above a threshold amplitude $A_o=1.4$ $(\hbar/ea_o)$, the charge order melts.

The Ni$_S$ magnetic moment remains negligible both before and after the light pulse; see Fig. \ref{fig:fig2}(b). As each Ni$_S$ site is
surrounded by opposite spin-polarized Ni$_L$ neighbors, the charge-transfer to Ni$_S$ during optical excitations is spin-compensated,
which explains the negligible net moment on Ni$_S$ in the photoexcited state. On the other hand, the Ni$_L$ magnetic moment reduces
with increasing light intensity, thus melting the original AFM spin-order.  Our results are consistent with a previous ultrafast experimental
observation of magnetic order melting using time-resolved XRD and magnetic scattering studies of optically excited RNiO$_3$
\cite{Caviglia2013,Stoica2022}.

\begin{figure}[tp!]
     \begin{center}
     \includegraphics[width=1.0\linewidth]{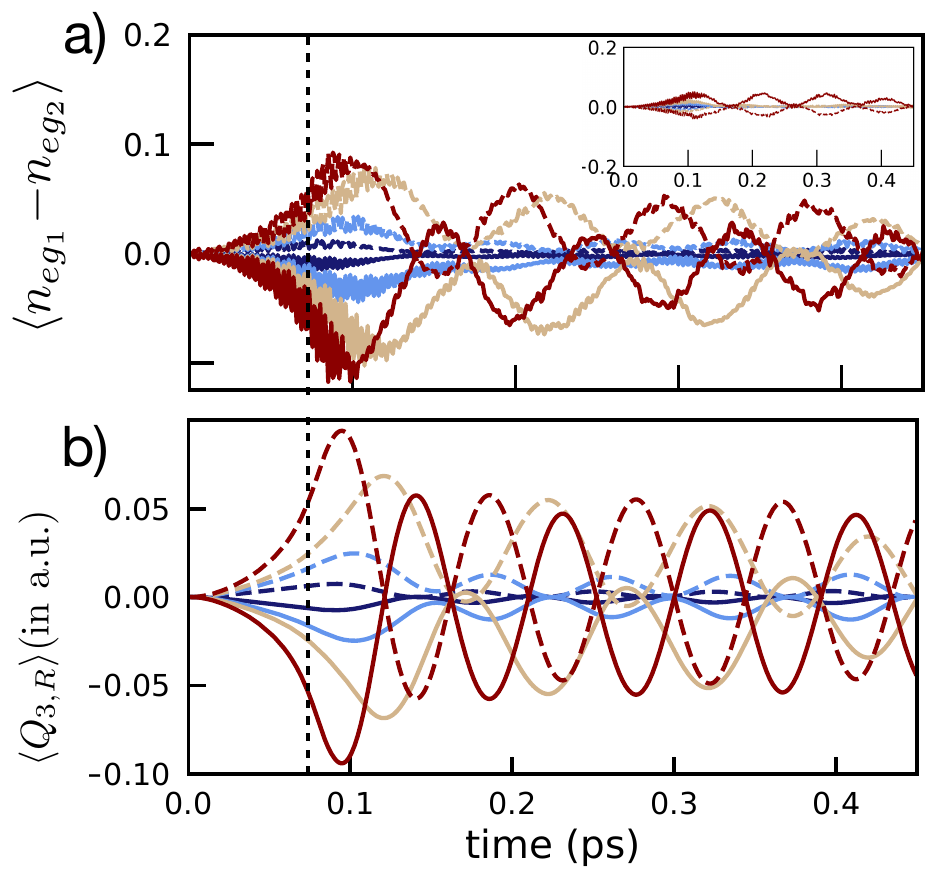}
\caption{  Time evolution of the average   
      the \(e_g\)-orbital occupancy difference \textbf{(a)} JT mode $\langle Q_{3,R}\rangle$  \textbf{(a)} at Ni\(_L\) with solid and Ni\(_S\) with
      dashed lines for several pump fluences, during and after photoexcitation with light polarized along \(\hat{\mathbf e}_z\). Inset (a): \(e_g\)-occupancy difference
      under \(\hat{\mathbf e}_x \pm \hat{\mathbf e}_y\) polarization. Colors indicate increasing pulse intensity as defined in Fig.~2.  }
     \label{fig:fig3}
     \end{center}
\end{figure}

\noindent
\textit{Orbital-polarization:} 
Interestingly, $e_g$-orbital polarization, absent in the ground state, emerges in the photoexcited state. Figure~\ref{fig:fig3}(a) shows the
time evolution of the orbital polarization at Ni sites for different light polarizations quantified by $\Delta n^{eg}_R{=} \sum_\sigma \rho_{\sigma\alpha_1,\sigma\alpha_1}{-}\rho_{\sigma\alpha_2,\sigma\alpha_2}$, 
where $\alpha_1{=}d_{x^2-y^2}$ and $\alpha_2{=}d_{3z^2-r^2}$. For isotropic excitation along $\hat{\mathbf e}_x {+} \hat{\mathbf e}_y {+} \hat{\mathbf e}_z$, $\Delta n^{eg}_R$
remains very small as shown in the inset of Figure \ref{fig:fig3}(a). In contrast, for $\hat{\mathbf e}_z$ polarization, a pronounced orbital
polarization develops and increases with light field strength. This behavior originates from polarization-dependent, orbital-selective dipole
transitions: light polarized along $\hat{\mathbf e}_z$ primarily couples to $d_{3z^2-r^2}$ orbitals, whereas in-plane polarization ($\hat{\mathbf e}_x{\pm}\hat{\mathbf e}_y$)
preferentially couples to $d_{x^2-y^2}$. As a result, the photoinduced charge transfer from Ni$_L$ to Ni$_S$ becomes orbital-selective, leading to an imbalance
in $e_g$-orbital occupations and a finite $\Delta n^{eg}_R$. In contrast, isotropic excitation maintains nearly balanced $e_g$-populations.

\begin{figure}[tp!]
     \begin{center}
     \includegraphics[width=1.0\linewidth]{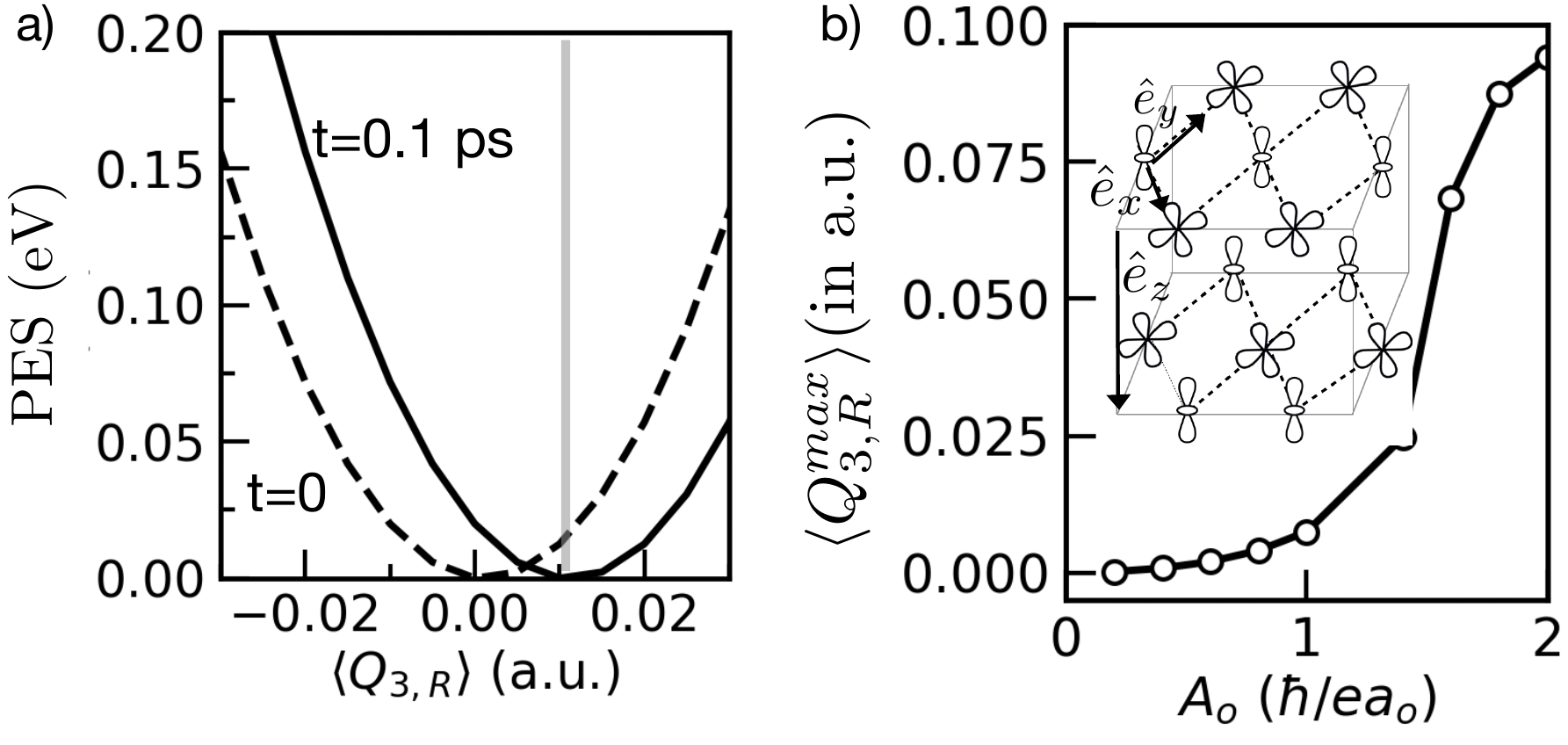}
     \caption{\textbf{(a)} Potential energy surface calculated as a function of JT mode $\langle Q_3\rangle$ for ground state (t=0) and
     for excited state at t=0.1 ps. The vertical line show the energy minima in the excited state shifted away from $Q_3=0$. \textbf{(b)} 
     Photoinduced long-range orbital order (inset) and maximum $\langle Q_{3,R}\rangle$ amplitude in the excited state as a function
     of the vector-potential amplitude \(A_0\). after photoexcitation with light polarized
     along \(\hat{\mathbf e}_z\).}
     \label{fig:fig4}
     \end{center}
\end{figure}

The resulting photoexcited state, characterized by a finite orbital population imbalance $\Delta n^{eg}_R$, 
becomes unstable toward JT distortions and lowers its energy by developing a finite lattice distortion, as shown in Fig.~\ref{fig:fig3}(b). 
This behavior can be understood from an effective time-dependent potential of the form
\begin{equation}
E(Q_{3,R},t) = \frac{1}{2}K Q_{3,R}^2 - g_{JT}\,Q_{3,R}\,\Delta n^{eg}_R(t),
\end{equation}
where the second term acts as a light-induced symmetry-breaking field. The excited-state PES, evaluated at a fixed
photoexcited electronic density, therefore exhibits a single-well profile with its minimum shifted to a finite value of $Q_3$,
see Fig.~\ref{fig:fig4} (a). The sign of $Q_3$ is determined by the polarization of light through selective orbital excitation. 
The linearly polarized photoexcitation biases the initial lattice dynamics toward a preferred JT distortion, consistent with
our ensemble Ehrenfest simulations (see Supplementary Material), where sampling over initial nuclear conditions yields a
finite ensemble-averaged order parameter $\langle Q_{3,R}(t)\rangle_{\mathrm{ens}} \neq 0$. 
The local orbital polarization in the photo-excited state forms a long-range orbital-order pattern, shown in
the inset of Fig.~\ref{fig:fig4} (b). The symmetry of the local JT distortions at Ni sites is consistent with the local
$e_g$-orbital polarization symmetry.  

For weak excitation, the lattice dynamics remain confined around this displaced minimum, resulting in small-amplitude
oscillations. However, for higher excitation strengths $A_o>$ 1.4 $(\hbar/ea_o)$, the lattice acquires sufficient kinetic
energy and, together with the time-dependent reduction of electronic bias $\Delta n^{eg}_R(t)$ (Fig.~\ref{fig:fig3}(a)),
allows $Q_{3,R}(t)$ to traverse through $0$ and explore both positive and negative distortions, as shown in Fig.~\ref{fig:fig3}(b).
The resulting large-amplitude coherent oscillations between $+Q_0$ and $-Q_0$ arise from the interplay of lattice inertia
and the time-dependent reshaping of the potential.

The maximum amplitude of the JT
mode $Q_{3,R}$ emerging in the excited state scales
nonlinearly with increasing vector-potential amplitude $A_0$, as shown in Fig.~\ref{fig:fig4}(b).

Our results demonstrate that light polarization can be used to selectively couple to specific $e_g$ orbitals in charge-ordered RNiO$_3$,
enabling orbital-selective $e_g \rightarrow e_g$ transitions and driving JT dynamics. The imbalance in $e_g$-orbital populations also leads
to a transient reduction of the effective Hund’s coupling $J$, which further enhances the susceptibility of the system to lattice distortions and
modifies the excited-state PES. Similar transient renormalizations of interaction parameters have been reported in correlated systems \cite{Baykusheva2022,Granas2022,Dejean2018}.

The photoinduced orbital polarization and JT dynamics predicted here can be probed using ultrafast pump-probe techniques with controlled
polarization, combined with time-resolved x-ray scattering \cite{Ehrke2011,Johnson2015,Beaud2007} and Ni $L$-edge resonant inelastic x-ray
scattering \cite{Dean2016,Mitrano2019,Cao2019}, which are sensitive to lattice symmetry breaking and orbital excitations, respectively.
Larger-bandwidth members of the RNiO$_3$ series are promising platforms to realize these effects. The oscillatory $\langle Q_{3,R}(t)\rangle_{\mathrm{ens}}$ 
would manifest as time-dependent superlattice peaks at $\mathbf{G} \pm \mathbf{Q}_{\mathrm{JT}}$. Although orbital order shares the same periodicity
as equilibrium charge order, distinct oscillation frequencies of the breathing and JT modes should allow their separation in the frequency domain.

In summary, using a time-dependent interacting multi-orbital tight-binding model, we show that polarization-selective optical excitation drives
 a nonthermal pathway to a photoexcited state with finite orbital polarization at Ni sites, which is absent in equilibrium. The light polarization
 controls coupling to specific $e_g$ orbitals, generating a transient orbital imbalance that is stabilized by coupling to JT modes, extending its
 lifetime beyond the pulse duration. These results demonstrate that tailored optical fields can activate otherwise inert orbital degrees of freedom
 and access hidden orbitally polarized phases, enabling ultrafast control of coupled charge, spin, and orbital dynamics.

\textit{Acknowledgments:} 
S.R, S.R.U.H., A.M.L., and T.O. are supported by the Computational Materials Sciences Program funded by the US Department of
Energy, Office of Science, Basic Energy Sciences, Materials Sciences and Engineering Division. This work is funded in part by the
Deutsche Forschungsgemeinschaft (DFG, German Research Foundation) 217133147/SFB1073, projects B03 and C03.

\bibliography{ref}
\newpage
\end{document}